\documentclass[a4paper,fleqn,usenatbib]{mnras}
\usepackage[utf8]{inputenc}

\usepackage{newtxtext,newtxmath}
\usepackage{multirow}

\usepackage[T1]{fontenc}
\usepackage{ae,aecompl}

\usepackage{graphicx}	
\usepackage{amsmath}	
\usepackage{amssymb}	

\newcommand{\flux}{erg s$^{-1}$ cm$^{-2}$}
\newcommand{\lum}{erg s$^{-1}$}	

\newcommand{\sxp}{{SXP\,4.78}}
\newcommand{\rxte}{\textit{RXTE}}
\newcommand{\chandra}{\textit{Chandra}}
\newcommand{\swift}{\textit{Swift}}
\newcommand{\swiftxrt}{\textit{Swift}/XRT}

\newcommand{\xmm}{\textit{XMM-Newton}}
\newcommand{\nicer}{\textrm{NICER}}
\newcommand{\nustar}{\textit{NuSTAR}}

\newcommand {\be}{\begin {equation}}
\newcommand {\ee}{\end {equation}}

\title[Magnetic field of \sxp]{Observational constraints on the magnetic field of the bright transient Be/X-ray pulsar \sxp}

\author[Semena A. et al.]
{Andrey N. Semena,$^{1}$\thanks{E-mail: san@iki.rssi.ru}
Alexander A. Lutovinov,$^{1}$
Ilya A. Mereminskiy,$^{1}$
\newauthor
Sergey S. Tsygankov,$^{2,1}$
Andrey E. Shtykovsky,$^{1}$
Sergey V. Molkov$^{1}$
and Juri Poutanen$^{2,1,3}$
\\
$^{1}$Space Research Institute of the Russian Academy of Sciences, Profsoyuznaya Str. 84/32, Moscow 117997, Russia\\
$^{2}$Department of Physics and Astronomy, FI-20014 University of Turku, Finland\\
$^{3}$Nordita, KTH Royal Institute of Technology and Stockholm University, Roslagstullsbacken 23, SE-10691 Stockholm, Sweden
}

\date{Accepted XXX. Received YYY; in original form ZZZ}

\pubyear{2019}

\begin{document}
\label{firstpage}
\pagerange{\pageref{firstpage}--\pageref{lastpage}}
\maketitle


\begin{abstract}
We report results of the spectral and timing analysis of the Be/X-ray pulsar SXP~4.78 using the data obtained during its recent outburst with \nustar, \swift, \chandra\ and \nicer\ observatories.
Using an overall evolution of the system luminosity, spectral analysis and variability power spectrum we obtain constraints on the neutron star magnetic field strength.
We found a rapid evolution of the variability power spectrum during the rise of the outburst, and absence of the significant changes during the flux decay.
Several low frequency quasi-periodic oscillation features are found to emerge on the different stages of the outburst, but no clear clues on their origin were found in the energy spectrum and overall flux behaviour.
We use several indirect methods to estimate the magnetic field strength on the neutron star surface and found that most
of them suggest magnetic field $B \lesssim 2 \times10^{12}$~G.
The strictest upper limit comes from the absence of the cyclotron absorption features in the energy spectra and suggests relatively weak magnetic field $B < 6 \times 10^{11}$~G.

\end{abstract}




\section{Introduction}

Discovery of pulsating ultra-luminous X-ray sources 
\citep{2014Natur.514..202B, 2016ApJ...831L..14F, 2017Sci...355..817I, 2017MNRAS.466L..48I,2019arXiv190604791R} unambiguously showed that neutron stars (NSs) can exceed their Eddington limit by orders of magnitudes. 
One of the necessary condition for existence of such a high luminosity was proposed to be a superstrong magnetar-like magnetic field which may significantly reduce the scattering cross-section   
\citep{2015MNRAS.454.2539M,2015AN....336..835T}. 
The magnetic field also regulates the mass accretion rate reaching the NS surface and therefore the limiting luminosity \citep{1982SvA....26...54L, 2019A&A...626A..18C}. 
Bright transient X-ray pulsars (XRPs) can be considered as a connecting link between a well-studied sample of moderately bright sources with known magnetic field and ultra-luminous X-ray pulsars. In order to reveal this possible link one need to determine magnetic field strength in the brightest representatives of the XRPs family.
 Do ultra-luminous X-ray pulsars form a new class of systems or they
are usual high mass X-ray binaries (HMXBs), occasionally accreting at the
super-Eddington rate, is still an open question.

To approach answering this question it is important to study usual XRPs at high luminosities. A significant fraction of such systems appear
to be binaries with Be star companion \citep{2015A&ARv..23....2W} at highly
elliptical orbits, referred to as Be X-ray binaries or BeXRBs. They 
are transient systems, demonstrating different types of outbursts \citep[see review by][]{2011Ap&SS.332....1R}. During so-called Type II outbursts these
systems often reach or even exceed the Eddington limit \citep[see,
e.g.,][]{2017A&A...605A..39T, 2018MNRAS.479L.134T}, which make them good
candidates to search connections with ULX pulsars and to study the accretion
at high rates.

\sxp\ (or XTE\,J0052$-$723) was discovered by the \rxte\ observatory during scanning
observations of the Small Magellanic Cloud in December 2000
\citep{2001IAUC.7562....1C}. These observations revealed a bright X-ray
source with coherent pulsations at $\sim4.782$ s and a double-peaked pulse
profile \citep{2003MNRAS.339..435L}. Four more outbursts from this source
were detected by \rxte\  \citep{2014MNRAS.437.3863K} during which its
luminosity reached $\sim 10^{38}$ \lum\ \citep{2003MNRAS.339..435L},
assuming the distance of 60.3 kpc. Taking into account such a transient
behaviour the source was suggested to be a BeXRB, although it was
impossible to achieve a precise localization with these scanning observations
and to determine its optical counterpart. There were several attempts to
estimate the magnetic field strength of \sxp, but an insufficient
sensitivity, low number of observations and neglect of the orbital motion
lead to a wide range of possible values of the magnetic filed in the range 
$10^{11}$--$4\times10^{13}$\,G \citep{2003MNRAS.339..435L,
2014MNRAS.437.3863K, 2015ApJ...813...91S}.

The source once again went into the outburst in November 2018 and was
detected by the \swift\ observatory \citep{2018ATel12209....1C}. A
precise localization of \sxp\ obtained with the XRT telescope on board of \swift\ allowed an identification of its optical counterpart -- the bright star [M2002]
SMC 20671. \citet{2019MNRAS.485.4617M} investigated optical spectra of this
star and deduced its class as B0.5-IV-V with a strong emission H${\alpha}$
line thereby confirming a Be-nature of the system. These authors inspected
also the long-term optical variability of the object with the OGLE project data and found no
clear signatures of the orbital period. The outburst was observed extensively
by many orbital X-ray observatories and telescopes providing a vast amount of data
in the broad energy band (0.2--78~keV). Based on the \nustar\ observation on 2018
Nov 15-17, \citet{2018ATel12234....1A} reported an evidence
for a cyclotron resonance scattering feature (CRSF) at $\sim10$\,keV, which is still unconfirmed.

In this paper, based on the data from \swift, \nicer, \nustar, \xmm, and \chandra\ observatories, we carried out comprehensive
timing and spectral analysis of the source properties during the 2018
super-Eddington Type II outburst with the main aim to constrain the
unknown strength of the magnetic field in this system. In
Section~\ref{sec:data} we describe the available data and their reduction. 
Section~\ref{sec:var} presents the results of the timing analysis and magnetic
field estimations from the propeller effect and power spectra. 
In Section~\ref{sec:spec} we investigate the X-ray spectrum of the source and present a 
search for the cyclotron line. The summary of results  obtained with different
methods is provided in Section~\ref{sec:concl}.

\section{Observations and data reduction}
\label{sec:data}	

After the detection of the new X-ray outburst of \sxp\ by the \swift\
observatory, it was frequently monitored with \nicer\ and \swiftxrt, counting
up 69 and 34 observations, correspondingly, distributed over 2018 Nov  -- 2019 Feb. Also, there were three long \nustar\ pointings near the outburst peak. At
the late stage of the outburst we initiated the \chandra\
Target-of-Opportunity observation (PI Lutovinov), which was performed on 2019 March 1. To estimate the lowest possible flux of \sxp, we used also archived
observations with the \xmm\ observatory, performed in 2007-2016.

Depending on the capabilities of instruments, these data were utilized for
spectral and/or timing analysis. Whenever energy spectra are obtained they
were grouped to have at least 30 counts per bin with \texttt{grppha} tool.
The final data analysis (timing and spectral) was performed with the \textsc{heasoft} 6.25 software package. All uncertainties are quoted at the $1\sigma$
confidence level, if not stated otherwise.

\subsection{\nustar\ data}

The \nustar\ space observatory consists of two identical X-ray telescope
modules, each equipped with independent mirror systems and focal plane
detector units, also referred to as FPMA and FPMB
\citep{2013ApJ...770..103H}. It provides X-ray imaging, spectroscopy and
timing in the energy range of 3--79~keV with the angular resolution of
18\arcsec\ (FWHM) and spectral resolution of 400~eV (FWHM) at 10~keV.

\begin{table}
\caption{\nustar\ observations of \sxp\ and measured pulse periods, and quasi simultaneous \swiftxrt\ observations.}\label{tab:obsnus}

	\centering
	\footnotesize

	\begin{tabular}{cccc}
		\hline\hline
		ObsID & Date & Exposure, ks & Period, s\\
		\hline
        \multicolumn{4}{c}{\nustar} \\
		30361003002 & 2018-11-15 & 74.5 & $4.781607\pm0.000005$ \\
		30361003004 & 2018-11-24 & 38.3 & $4.781671\substack{+0.000001\\-0.000005}$\\
		30361003006 & 2018-11-27 & 73.2 & $4.781656\pm0.000001$ \\
        \multicolumn{4}{c}{\swift}\\
        00010977003 & 2018-11-16 & 4.9  &  \\
        00010977004 & 2018-11-25 & 6.9  &  \\
        00031998011 & 2018-11-27 & 5.9  &  \\

		\hline
	\end{tabular}
\end{table}

\nustar\ performed three observations of SXP~4.78 during the outburst, all
were held near its maximum (see Table ~\ref{tab:obsnus}). Hereafter we refer
to these observations as first, second and third ones based on their
chronological order.
To process the \nustar\ data we used the standard \textsc{nustardas} 1.8.0
software as distributed with the \textsc{heasoft} 6.25 package and the
\textsc{CALDB} version 20181030. The standard \textsc{lcmath} tool was used
to combine the light curves of the \nustar\ modules to improve statistics for
the timing analysis. Source data were extracted from a circular region with
radius of 90\arcsec, centered at its position. Background data were
extracted using a region of an equivalent area away from the source position.
Observational data have no signs of a contamination by a stray-light or ghost
rays.

\subsection{\nicer\ data}

The Neutron star Interior Composition Explorer (\nicer)  is an International Space Station payload consisting of 56 "concentrator" optics above silicon drift detectors operating in the 0.2--12~keV
energy band \citep{2016SPIE.9905E..1HG}.  
It provides more than 2000 cm$^2$ effective area in the soft X-rays, excellent timing capabilities and CCD-like spectral resolution. 
Due to the availability of the \nustar\ and \swiftxrt\ data, well calibrated 
and suitable for the broadband spectral analysis, we used the \nicer\ data
only for the analysis of the temporal properties of the source. 


The data reduction was performed with the standard \textsc{nicerdas} software
(V005). We applied a barycentric correction to all cleaned event lists. As it
was discussed by \citet{2016SPIE.9905E..1IP} so-called 
`hard tails',
produced by the high-energy particles, can be observed in the \nicer\ 
energy spectra. This component of the spectrum introduces an additional
stochastic noise which has a complex shape in the power spectra \citep[see,
e.g.][]{2018MNRAS.481.1658S}. In order to reduce the impact of this noise in
our data we applied additional filtering criterion to the \nicer\ 
data, following \citet{2018ApJ...860L...9B}. Initially, we build a light
curve in the 1200--1500 PI channels range (corresponding to 12--15~keV energy
band, where the instrument has negligible effective area). After that new
good time intervals, including only those part of the light curve with the
count rate below 1 count per second, were created. Note, that this procedure
usually leads to the loss of $\sim$10 per cent of the exposure time in each
individual observation.

The source was observed 69 times in the period from 2018 November 10 to 2019
March 12 (ObsIDs 12004101[01--69]). For our analysis we used all these
data, excluding observations with ObsIDs 1200410110 and 1200410141 for which
we could not make the barycentric correction. After the filtering we
obtained 90.5 ks of the total exposure. For the following analysis we use events
registered in 50--800 PI channels corresponding to the 0.5--8~keV energy band.

\begin{figure*}
    \centering
    \includegraphics[width=\textwidth]{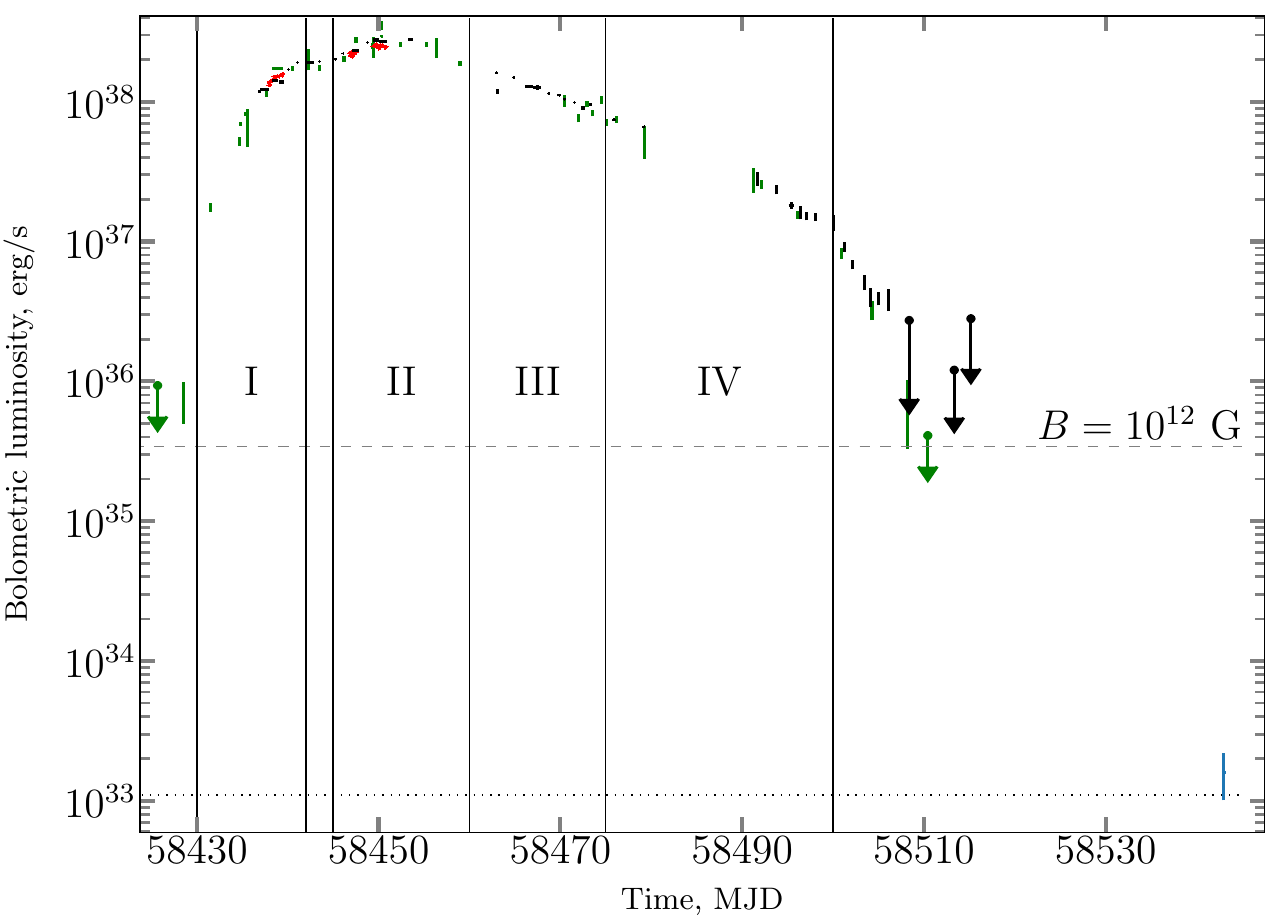}
\caption{Evolution of the \sxp\ bolometric luminosity during the outburst, as
measured by \swiftxrt\ (green points with error bar), \nustar\ (red) and \nicer\ (black). 
The \chandra\ observations in the quiescent state is
shown by the blue error bar. 
The dashed horizontal line represents the critical luminosity at which the source
should switch to the `propeller mode' for the magnetic field strength of
$10^{12}$~G. The dotted horizontal line demonstrates the upper limit for the
source luminosity derived from \xmm\ data obtained in 2009 (ObsID.
0601210701).} \label{fig:outburst_lcurve}
\end{figure*}

\subsection{\swift\ data}

The \textit{Neil Gehrels Swift Observatory} \citep{2004ApJ...611.1005G} carries several instruments including the glazing mirror X-ray telescope (XRT) allowing to perform imaging, spectral and timing analysis in soft
X-rays \citep{2000SPIE.4140...64B}. The telescope can operate in the imaging
and timing modes. \sxp\ was observed 35 times since its first detection on
2018 November 9  till the last detection on 2019 February 24. A median exposure of the
single observation was about 1\,ks. Only five of these observations were
performed in windowed timing mode, therefore we used \swiftxrt\ data only for
spectral analysis. The spectra were produced by the online tool
\citep{2009MNRAS.397.1177E}. 
The data analysis was restricted to the 0.8--9~keV energy band.

\subsection{\chandra\ data}

At the late stage, when the source became too weak for \swiftxrt, we
employed the \chandra\ observatory to determine the source state and measure its
flux. The observation was carried out on 2019 Mar 1 (ObsId. 22095). The
standard {\sc ciao} package with latest calibration files was used to analyse
these data. First, the obtained data were reprocessed to produce the
sky map in the 0.5--7 keV energy band. No significant background variations
over the course of observation were found. Usage of the {\sc wavdetect}
procedure yields a 2.7$\sigma$ detection of the source at 0.44\arcsec\ away
from the optical counterpart of \sxp. This offset is consistent both with
measured uncertainties on the source position and the typical \chandra\ astrometric precision.

\subsection{\xmm\ data}

The field of \sxp\ was observed with the \xmm\ observatory three
times in the past, but the source was never detected. Nevertheless the
knowledge of the source flux in the quiescent state is important for
understanding the presence of the propeller regime and estimation of the
magnetic field. Therefore we used the {\sc FLIX} service \citep{2007A&A...469...27C}\footnote{\url{https://www.ledas.ac.uk/flix/flix_dr7.html}} 
to estimate the $4\sigma$ upper limits on the source flux in these
observations. The obtained limits are presented in Table\,\ref{tab:obsxmm}.

\begin{table}
\caption{\xmm\ observations of \sxp}\label{tab:obsxmm}
	\centering
	\footnotesize
	\begin{tabular}{cccc}
		\hline\hline
		ObsID & Date & Exposure & Flux upper limit  \\
		      &      & ks       & erg~s$^{-1}$~cm${^-2}$ \\
		\hline
		0500980101 & 2007-06-23 & 24 & $7.0\times10^{-15}$ \\
		0601210701 & 2009-09-27 & 37 & $2.5\times10^{-15}$\\
		0793182901 & 2016-10-14 & 35 & $3.2\times10^{-15}$ \\
		\hline
	\end{tabular}
\end{table}

\subsection{Bolometric correction}

The 2018 outburst of \sxp\ was well covered by monitoring observations with
\swiftxrt\ and \nicer\ instruments, allowing us for the first time to examine
its profile in details from the rise to the decline. It is necessary to note that
these monitoring programs were performed in the X-ray range with energies not exceeding 10--12 keV,
but for estimations of physical properties of the source (e.g. the magnetic
field) we need to know its bolometric luminosity (we consider it as a
luminosity in the 0.1--100 keV energy band). To reconstruct the broad-band
spectrum of the source and to calculate the bolometric correction we used three
\nustar\ (and nearly simultaneous \swiftxrt) observations as reference points
(see Section\,\ref{sec:spec} for details). It was found that the bolometric correction from the luminosity in the 0.5--8 keV energy band slightly varied as $K_{\rm bol}\simeq(2.2-2.4)$, but the
spectral shape remained approximately the same. Therefore,  in the following as a bolometric correction we adopted $K_{\rm bol}=2.3$.

\section{Timing analysis}
\label{sec:var}

As it was mentioned in the introduction,  the \rxte\ data
allowed measurements of some timing properties of \sxp\ (pulsations, pulse
profiles, etc.). However, the small number of the available observations prevented in depth timing
studies of the source, especially at low luminosities. In this section, we
perform the timing analysis of newly available data in the X-ray band,
accumulated with \nustar\ and \nicer\ during the recent outburst
of the source. \swiftxrt\ data were used also in order to trace the overall
evolution of the source luminosity during the outburst.

\subsection{Long term temporal behaviour}

The overall outburst profile is shown in Fig.~\ref{fig:outburst_lcurve}. The
first detection corresponds to 2018 Nov 9  (MJD\,58431) when the source was
revealed by \swiftxrt\ with the bolometric luminosity of $L_{\rm bol}\simeq 7
\times 10^{35}$\lum. In the beginning of the outburst the source underwent
fast rise, increasing its luminosity by a hundred times over nine days and
reaching $10^{38}$\lum. Then the growth rate was decreased
and in the consecutive dozen days the luminosity has increased from $10^{38}$ to $2.5\times10^{38}$~\lum. After
reaching the maximum a gradual decay started, lasting for approximately 50
days. As soon as the luminosity has decreased down to $L_{\rm bol}\simeq
2\times 10^{37}$~\lum, a nearly power-law decay switched to a faster decay. 
Eventually, the source was not detected by \swiftxrt\ with the upper limit of $L_{\rm
bol} < 4.1 \times 10^{35}$~\lum ($2\sigma$) on 2019 Jan 27  (MJD\,58510)
during a long ($\simeq4$\,ks) observation.

Such a behaviour is typical for fast rotating NSs in transient BeXRBs and indicates the transition of the system to the
propeller regime \citep{1975A&A....39..185I}.
Therefore, after the beginning of the rapid flux decay we initiated the \chandra\ ToO observation with the main aim to observe
the source in the quiescence state. 
The observation was performed on 2019 Mar 1. 
Despite only the marginal detection of the source it still had an important scientific significance. 
The source count rate in the 0.5--7 keV energy band
was estimated with the {\sc srcflux} tool of the {\sc ciao} package as
$3.7\times10^{-4}$~s$^{-1}$ with the 90 per cent confidence interval being
(1.5--7.4)$\times10^{-4}$~s$^{-1}$. Assuming a blackbody spectrum with the
temperature of $kT=0.5$~keV, that is expected and observed for sources in the propeller regime \citep[see, e.g.,][]{2016A&A...593A..16T,
2016MNRAS.463L..46W}, and the absorption column of $2\times10^{21}$~cm$^{-2}$
\citep{2005A&A...440..775K}, we estimated an unabsorbed source flux of
$3.7\times10^{-15}$\flux\ and the corresponding intrinsic luminosity of
$L_{q}\simeq1.6\times10^{33}$\lum\ for the distance of 60.3 kpc. Such a
luminosity agrees well with usual values for HMXBs in the propeller state
\citep{2016A&A...593A..16T, 2016MNRAS.463L..46W, 2017MNRAS.470..126T, 2017ApJ...834..209L,
2019MNRAS.485..770L}.

As it was mentioned above, the source was traced
until the \swiftxrt\ non-detection at the luminosity of $L_{\rm bol} \lesssim
4.1 \times 10^{35}$~\lum\ (Fig.~\ref{fig:outburst_lcurve}). 
Using this value as the upper limit on the critical luminosity of the transition to the propeller
regime we can constrain the NS magnetic field of \sxp\ as
\citep[see, for example,][]{2016A&A...593A..16T}: 
\be 
B_{\rm 12} \simeq 0.5
L_{\rm 37}^{1/2} P^{7/6} M_{\rm 1.4}^{1/3} R_{6}^{-5/2} k^{-7/4}, 
\ee 
where
$B_{\rm 12}$ is the magnetic field strength in 10$^{12}$~G, $L_{\rm 37}$ is
the luminosity in $10^{37}$\lum, $P$ is the spin period in seconds, $M_{\rm
1.4}$ is the NS mass in units of 1.4M$_\odot$, 
 $R_{6}$ is the NS radius in units of $10^6$\,cm, and $k$ is the ratio of the magnetospheric radius to
the Alfv\'en one, usually taken as 0.5 \citep{2018A&A...610A..46C}.
Substituting measured values for the source luminosity and the pulse period, and
using current results of the NS radius estimations $R_{6}=1.2$
\citep{2017MNRAS.472.3905S, 2017MNRAS.466..906S,Nattila.etal:17} we can
estimate an upper limit on the magnetic field strength as $B \lesssim
1.5\times10^{12}$\,G.

\begin{figure*}
        \includegraphics[width=\textwidth]{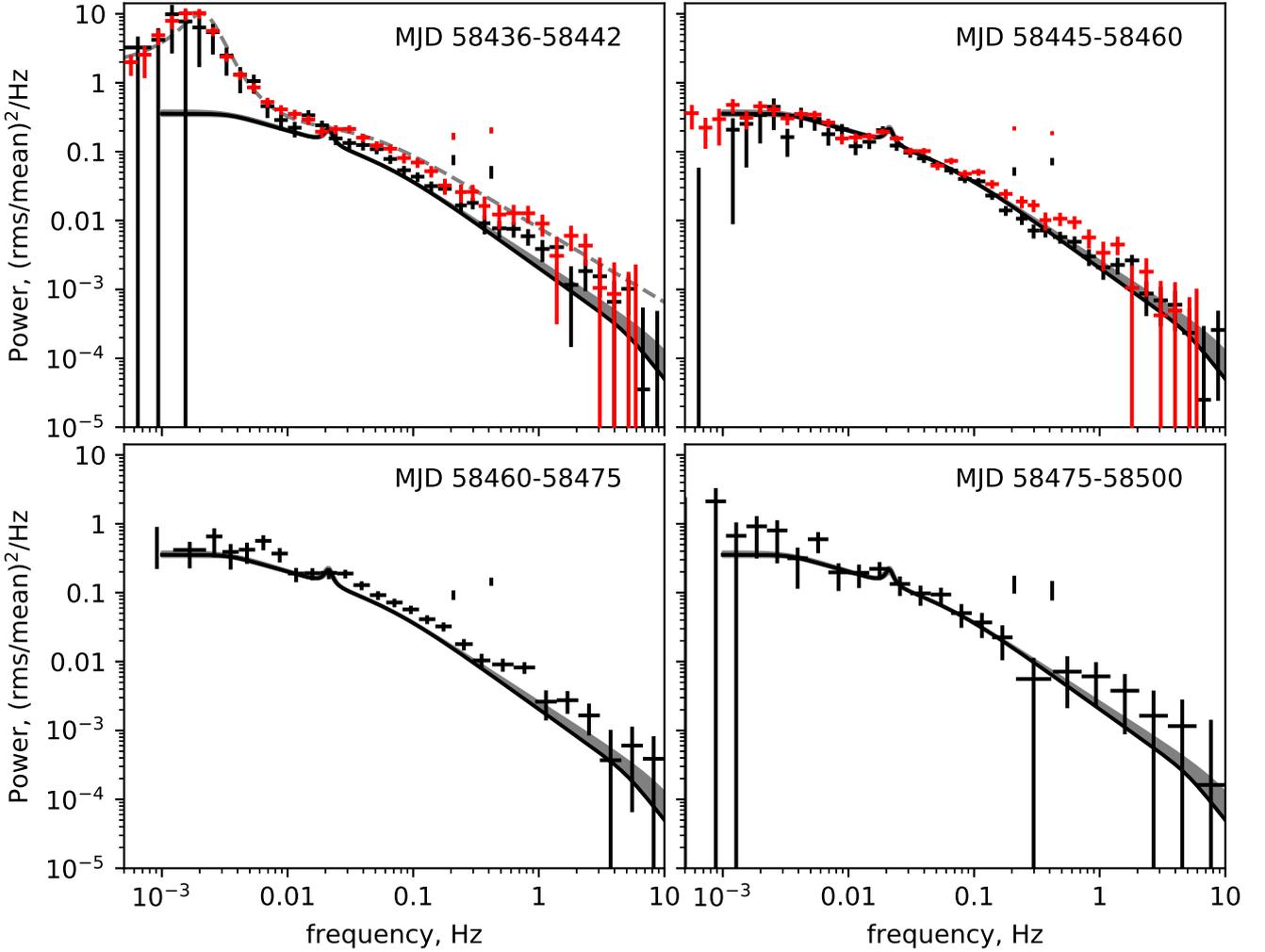}
\caption{Power spectra of \sxp, obtained with \nicer\ (black crosses) and
\nustar\ (red crosses) in four epochs of the outburst (see text and
Fig.\,\ref{fig:outburst_lcurve}). The black line represents the broken
power-law model, which fits the  \nicer\ data in the epoch II (top right
panel). The grey dashed line on the top left panel shows the  band-limited noise model with an additional Lorentzian introduced to describe the low-frequency QPO.}
\label{fig:lf_qpo}
\end{figure*}

\subsection{Power spectrum}

\begin{figure}
    \centering
    \includegraphics[width=\columnwidth]{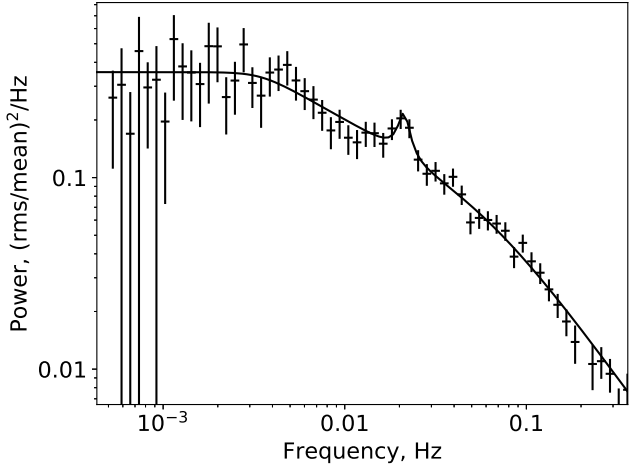}
    \caption{Power spectrum of \sxp\ at low frequencies obtained with \nicer\
    and \nustar\ in the epoch II (MJD 58445--58460).}
    \label{fig:ps_zoom}
\end{figure}

Power spectra of accreting systems can provide important information about
the geometry and properties of the accretion disc. 
In the broad frequency range, the power spectra can be described by a sum of several power-law or broad Lorentzian components \citep{2008A&A...489..725R}. 
At low frequencies, the power spectrum is usually a white noise $P(f)=const$, transiting to the $P(f)\propto f^{-1}$ at higher frequencies.
The $P(f)\propto f^{-1}$ segment of the power spectrum is thought to be produced by perturbations propagating in the accretion disc \citep{1997MNRAS.292..679L}.
Above a certain break frequency it transits to even steeper power law $P(f)\propto f^{-2}$.
\citet{2009A&A...507.1211R} suggested that this break frequency corresponds to a Keplerian frequency at the inner edge of the magnetically
truncated accretion disc. 
Thus the detection of such a break in the power spectrum of X-ray pulsars and its evolution with the source luminosity can
give an independent estimate of the magnetic field strength in the system \citep{2009A&A...507.1211R}. 

In the following, we investigate the variability properties of \sxp\ in terms of
its power spectrum and describe it with a combination of broad noise
components and coherent pulsations. Firstly, we analysed power spectra of each
separate \nustar\ and \nicer\ observation in the 3--78 and 0.5--8~keV
energy bands, respectively. We found, that during first twelve days of the
outburst the power spectrum had a form of a broken power-law continuum
(usually referred to as band-limited noise) and a broad hump-like
feature at low frequencies ($5\times10^{-4}$--$5\times10^{-3}$~Hz). The
presence of the excessive low frequency variability was reported earlier by
\citet{2018ATel12219....1G} based on the first portion of the \nicer\ data.
Interestingly, this feature disappears as the outburst evolves and the
luminosity increases. Therefore, in order to measure the power spectrum of
\sxp\ with a good accuracy and to trace its evolution over the outburst, we split all available data into four epochs
(Fig~\ref{fig:outburst_lcurve}): (I) beginning of the outburst (MJD
58436--58442), (II) near the outburst maximum (MJD 58445--58460), (III) slow
decay (MJD 58460--58475) and (IV) exponential decay (MJD 58475--58500), and
reconstructed power spectra for each of them (Fig.\,\ref{fig:lf_qpo}).

First, it should be noted that the power spectra obtained from \nustar\ and
\nicer\ data are in good agreement with each other for epochs I and II.
Second, the power spectra in epochs II, III and IV can be described with a
single zero-centered Lorentzian, representing a broad noise component, with
narrow spikes near the pulse frequency and its harmonic
(Fig.\,\ref{fig:lf_qpo}). The low-frequency broad noise does not change its
shape and amplitude significantly from epoch to epoch in spite of two orders of magnitude variations in the source
flux. The overall amplitude of this noise
component in both \nustar\ and \nicer\ energy bands was $ 14\pm1$ per cent (rms/mean) and the
low-frequency break of the component was measured at $F_{\rm lfbr}=
(3.5\pm0.7)\times10^{-3}$\,Hz. 

As noted above, in the power spectrum of epoch I, in addition to the band-limited noise component, the broad hump-like feature at low frequencies is
also significantly detected (Fig.\,\ref{fig:lf_qpo}). The overall power 
by this excessive variation is $15\pm2$ per cent (rms/mean). Similar features were observed in
a number of BeXRBs at different stages of outbursts
\citep[see][]{2008A&A...489..725R} and are known as milliherz quasi-periodic oscillations (mQPO).
\citet{2002ApJ...565.1134S} proposed a model of the accretion disc wrapped
under the influence of the NS magnetic torque to describe such mQPOs. Authors
noted, that the mQPO frequency depends in a complex way on the accretion
disc parameters which are hard to determine from the observations.
Fortunately, among the systems investigated by \citet{2002ApJ...565.1134S}
there is another BeXRB pulsar 4U\,0115+63, which demonstrated mQPO at 2~mHz
when the system has a similar to \sxp\ luminosity ($\approx
9\times10^{37}$\lum). This pulsar has a pulse period of 3.61~s and known
magnetic field strength ($B\approx1.3\times10^{12}$~G) due to the cyclotron
resonance features found in its spectrum \citep{1983ApJ...270..711W}. Using
a general dependence of the mQPO frequency on the magnetic moment derived in the
model of \citet{2002ApJ...565.1134S} $\nu_{\rm QPO}\propto \mu_{30}^{-0.81}$  we
can roughly estimate the magnetic field in \sxp\ as $B\approx
2\times10^{12}$~G.

All power spectra of \sxp\ at high frequencies approximately follow a
power law (Fig.\,\ref{fig:lf_qpo}) without an obvious presence of the high
frequency break, which can be associated with the accretion disc inner edge.
If the propagating fluctuations theory is valid, such a break should be located
at lower frequencies when BeXRBs are less luminous
\citep{2009A&A...507.1211R}.
However, it is easy to show, that the
significance of the break detection is larger when the system is bright,
because the break frequency decreases with the luminosity slower than the Poisson noise amplitude. Thus, for the power spectrum obtained in the brightest
epoch (II) we can constrain $F_{\rm br} > 5$~Hz, placing a lower limit
on the accretion disc inner radius, and therefore, an upper limit on the
magnetic field strength $B < 10^{12}$~G.

Finally, in order to investigate the power spectrum of \sxp\ in details we
used epoch II, which contain two long \nustar\ observations. A large number of
photons allowed us to build the power spectrum with the higher resolution
(Fig.~\ref{fig:ps_zoom}) and found that in line with other components the
spectrum has another QPO with the centroid at $(2.12\pm0.06)\times10^{-2}$~Hz
and relative rms of $1.6\pm0.2$ per cent (rms/mean). Note that a similar QPO was observed in
Cen\,X-3 \citep{1990A&A...230..103B}, but no model, describing its origin was
presented.

\subsection{Pulsations}

A detailed timing analysis is out of scopes of this paper, therefore we
limited ourselves here to only some general information about the temporal
properties of the source, derived from the \nustar\ data. First of all,
pulsations were clearly detected in all three observations and the pulse
period was determined with a high accuracy (see Table\,\ref{tab:obsnus}).
Although significant changes of the pulse period are observed, we cannot draw
any conclusions about the angular momentum transfer (which also can be used
for the magnetic filed estimations, see, e.g., \citealt{2014MNRAS.437.3863K})
in this system because of lack of orbital ephemeris.

\begin{figure}
    \includegraphics[width=\columnwidth]{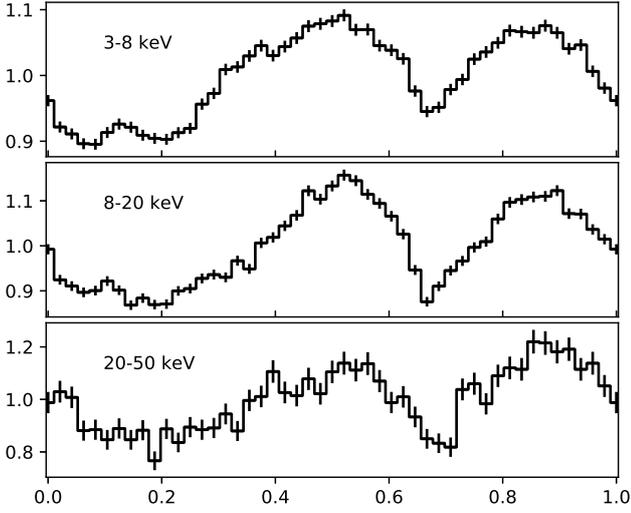}
    \caption{Pulse profile of \sxp\ in 3--8, 8--20, and 20--50~keV energy bands according to the \nustar\ data.}
    \label{fig:pprofile}
\end{figure}

Previously \citet{2003MNRAS.339..435L} reported a double-peaked pulse profile
in the 3--10 keV band. Here we used \nustar\ data to study the dependence of the pulse profile
and the pulse fraction with energy. The pulse profile has a
double-peaked shape, which does not change strongly with energy. We also
did not found any significant evolution of the pulse profile between three
\nustar\ observations, therefore only data from the brightest \nustar\
observation (ObsID\,30361003006) are presented in Fig.~\ref{fig:pprofile}.

\begin{figure}
    \centering
    \includegraphics[width=\columnwidth]{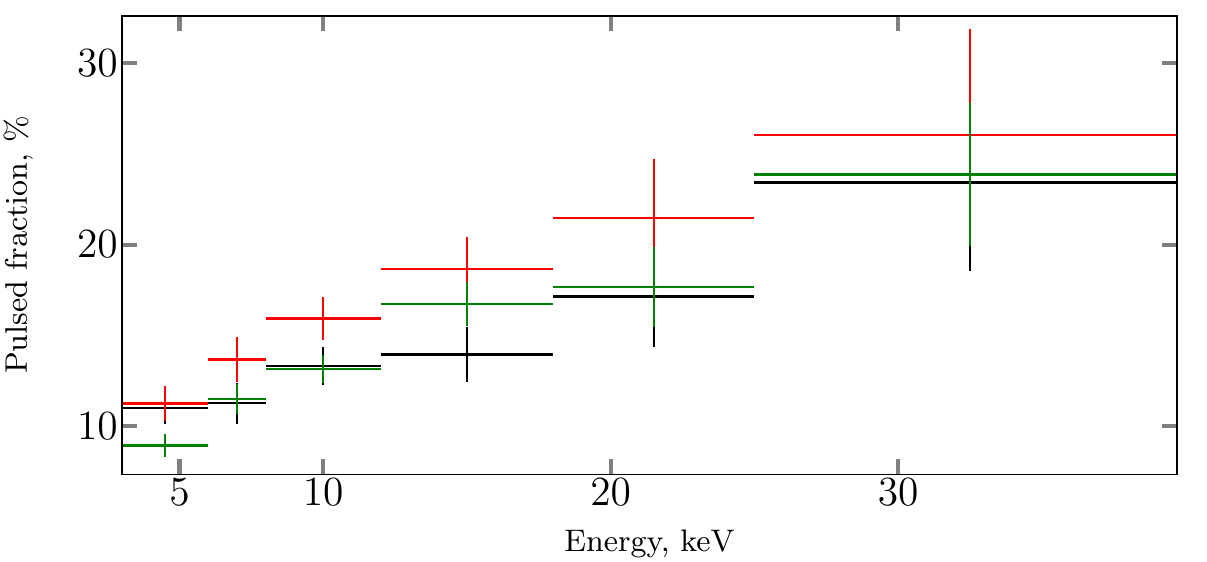}
    \caption{Dependence of the pulsed fraction on energy as measured
    by the \nustar\ observatory. Observations with ObsID 30361003002,
    30361003004 and 30361003006 are shown with the black, red and green crosses,
    respectively.} \label{fig:pulsed_fraction}
\end{figure}

The pulsed fraction, calculated as $PF=({F_{\rm max} - F_{\rm min}})/({F_{\rm
max} + F_{\rm min}})$, where $F_{\rm max}$ and $F_{\rm min}$ are the maximum
and minimum count rates in the profile, respectively, grows with the energy,
that is typical for bright X-ray pulsars \citep{2009AstL...35..433L}, and
does not evolve significantly from one \nustar\ observation to another
(Fig.\,\ref{fig:pulsed_fraction}). No additional peculiarities, which
previously were detected for several XRPs near the cyclotron line
energy \citep{2007AstL...33..368T,2009A&A...498..825F, 2009AstL...35..433L},
were found in the pulsed fraction behaviour of \sxp.

\section{Spectral analysis}
\label{sec:spec}

The spectral analysis is by far the most reliable method to estimate the
magnetic field strength through the detection of cyclotron absorption lines
in spectra of XRPs \citep[see][for the recent
review]{2019A&A...622A..61S}. Soon after the discovery of \sxp,
\citet{2003MNRAS.339..435L} investigated its spectrum with \rxte/PCA and
found that it can be well described with the bremsstrahlung emission without
any cyclotron absorption features. Recently, a preliminary spectral analysis
performed over the first \nustar\ observation (ObsID\,30361003002) revealed
an evidence for the possible absorption feature at $\sim10$\,keV
\citep{2018ATel12234....1A}. Assuming that this spectral feature may appear
to be a cyclotron absorption line, authors estimated the magnetic field of
the object as $B\approx0.9\times10^{12}$~G.

In this section a detailed analysis of the \sxp\ X-ray spectrum was performed
to examine the possible presence of the cyclotron absorption line and to
approximate the source spectrum in the broad energy range in the best way.

\subsection{Spectra of \sxp, measured with \nustar}
\label{sec:nuspec}

\begin{figure}
\centering
\includegraphics[width=\columnwidth]{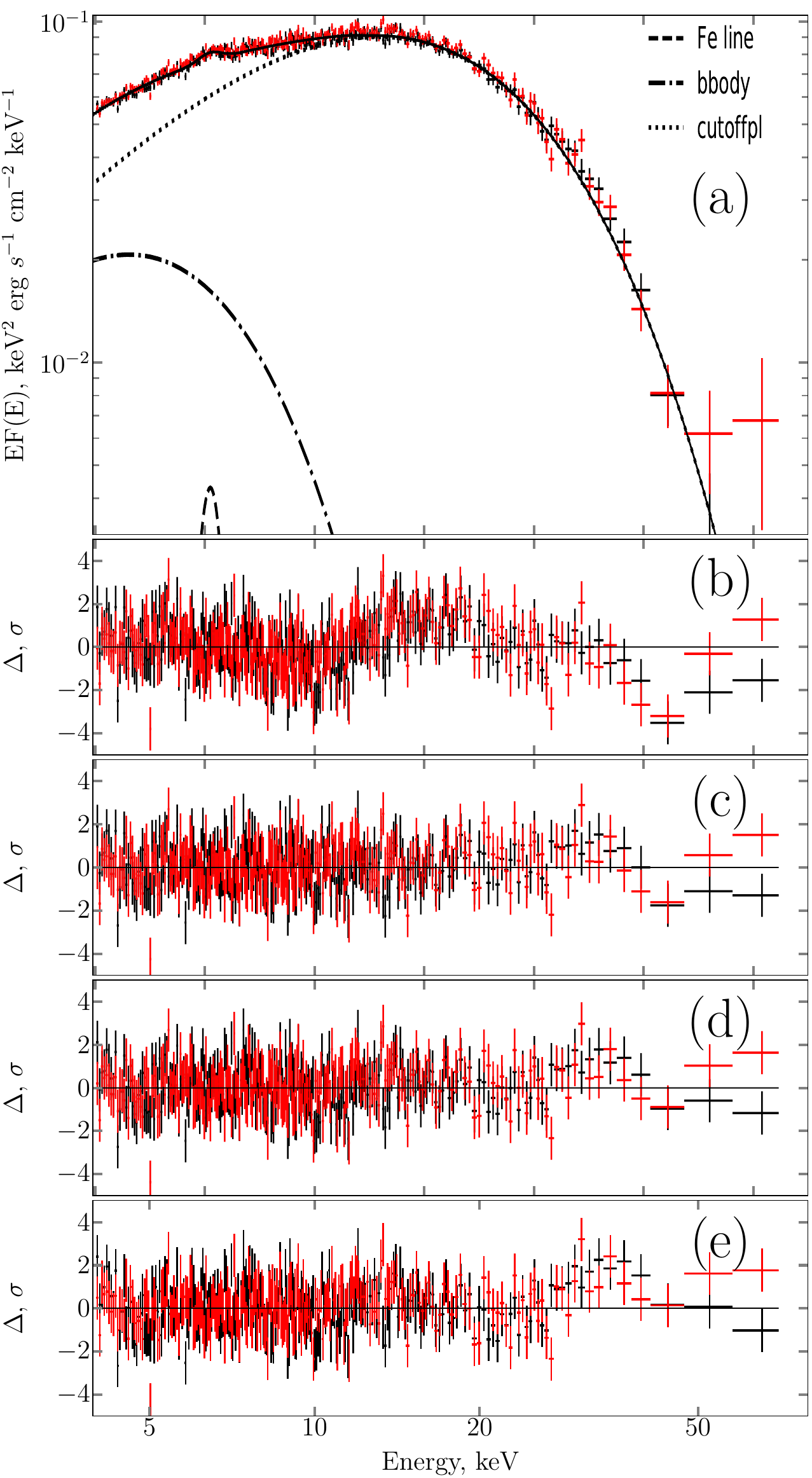} 
\caption{(a) Energy spectrum of \sxp\ obtained with the {\it NuSTAR}
observatory (ObsID\,30361003002). The black and red crosses correspond to data
from FPMA and FPMB, respectively. The black line shows the best-fitting model.
(b-e) Residuals from different models (see text for details). The spectrum
and residuals were rebinned in the figure to make deviations more apparent.}
\label{fig:cutoffpl_cyc_line}
\end{figure}

The spectrum of \sxp\ is typical for accreting XRPs and demonstrates
an exponential cutoff at high energies (Fig.\,\ref{fig:cutoffpl_cyc_line}a).
Therefore, as a first step, we attempted to approximate spectra 
obtained with the \nustar\ observatory, by the power-law model with an
exponential cutoff at high energies ({\sc cutoffpl} in the {\sc xspec}
package) and additional iron emission line at 6.4 keV in the form of the Gaussian. Note, that the same model was used by \citet{2018ATel12234....1A}
for the analysis of the first \nustar\ observation (ObsID\,30361003002). In agreement with these authors, we found that this model does not  describe well the
source spectrum. In particular, the residuals demonstrate a wave-like structure
with some deficit of photons around 10 keV
(Fig.\,\ref{fig:cutoffpl_cyc_line}b). A formal inclusion to the model of an
absorption component in the form of the {\sc gabc} model at the energy of
$E_{\rm cyc} = 9.78\pm0.15$~keV, width of $2.41\pm0.29$ keV and depth of
$\tau_{\rm cyc}=0.083_{-0.014}^{+0.017}$ improves the fit and flattens the residuals
(Fig.\,\ref{fig:cutoffpl_cyc_line}c). Other parameters of the best-fitting model
are the photon index $\Gamma=0.55\pm0.02$, the cutoff energy
$E_{\rm cut} = 7.80\pm0.12$~keV, and the iron line energy  $E_{\rm Fe} =
6.38\pm0.07$~keV; the iron line width was fixed at 0.3 keV, the value obtained by fitting the spectrum obtained during the third \nustar\ observation (with largest
photon statistics) with this parameter free.

At the same time, it is well known that the XRP spectra have a more
complex continuum shape than the power law with an exponential cutoff and
often several components are required to describe them \citep{2007ApJ...654..435B, 2012A&A...540L...1D, 2012MNRAS.421.2407T,2012A&A...538A..67F}. 
We found that this fully applies to the object under study. In particular, an addition to the {\sc cutoffpl} model of the thermal component in the form of the blackbody model with the temperature of $kT_{\rm BB}\simeq1.17$\,keV improves the fit quality similarly to adding the cyclotron line (Fig.\,\ref{fig:cutoffpl_cyc_line}d). 
Moreover, using another continuum model in the form of the comptonized emission ({\sc comptt} model in the {\sc xspec} package) allows us to describe the source spectrum much better that with the {\sc cutoffpl} model without a need to include the cyclotron absorption line (Fig.\,\ref{fig:cutoffpl_cyc_line}e).

The same analysis was carried out for two other \nustar\ observations of \sxp,
and similar results were obtained. Using of the {\sc cutoffpl} model for
the description of the continuum lead again to the wave-like structures in
the residuals, which can be compensated by addition of the broad absorption features at energy $E_{\rm cyc}\simeq11$ and 12.3~keV with depths of $\tau_{\rm cyc} \simeq 0.065$ and 0.14 for observations 30361003004 and 30361003006, respectively. Similarly to the  previous case, addition of the blackbody component with the
temperature $kT\simeq1.3$keV or using the {\sc comptt} continuum model allows us to
improve both fits without a need for the cyclotron absorption line.
  
Summarizing all above we can conclude the relative broadness of the possible
absorption feature, its small depth and proximity to the power-law cutoff
energy as well as the better approximation of spectra with other models
suggests the complexity of the spectral continuum in \sxp, rather than requires the presence of the cyclotron absorption line. 

\subsection{Broadband phase-averaged spectra}
\label{sec:avgspec}

To reconstruct the broadband spectra of \sxp\ we used all three \nustar\
observations with overlapping the data obtained with \swiftxrt\ (see Table~\ref{tab:obsnus}).
To take into account the uncertainty in the instrument calibrations, cross-calibration constants between them were included in all
spectral models (the $C_{\rm FPMB}$ and $C_{\rm XRT}$ constants correspond to the
cross-calibrations of the FPMB module and the XRT telescope to the FPMA
module, respectively). Given the lack of sensitive observations at lower
energies we fixed the value of the interstellar absorption column at
$2\times10^{21}$ cm$^{-2}$, which is typical in the direction to the Small 
Magellanic Cloud  \citep{2005A&A...440..775K}.

\begin{figure}
  \centering
  \includegraphics[width=\columnwidth]{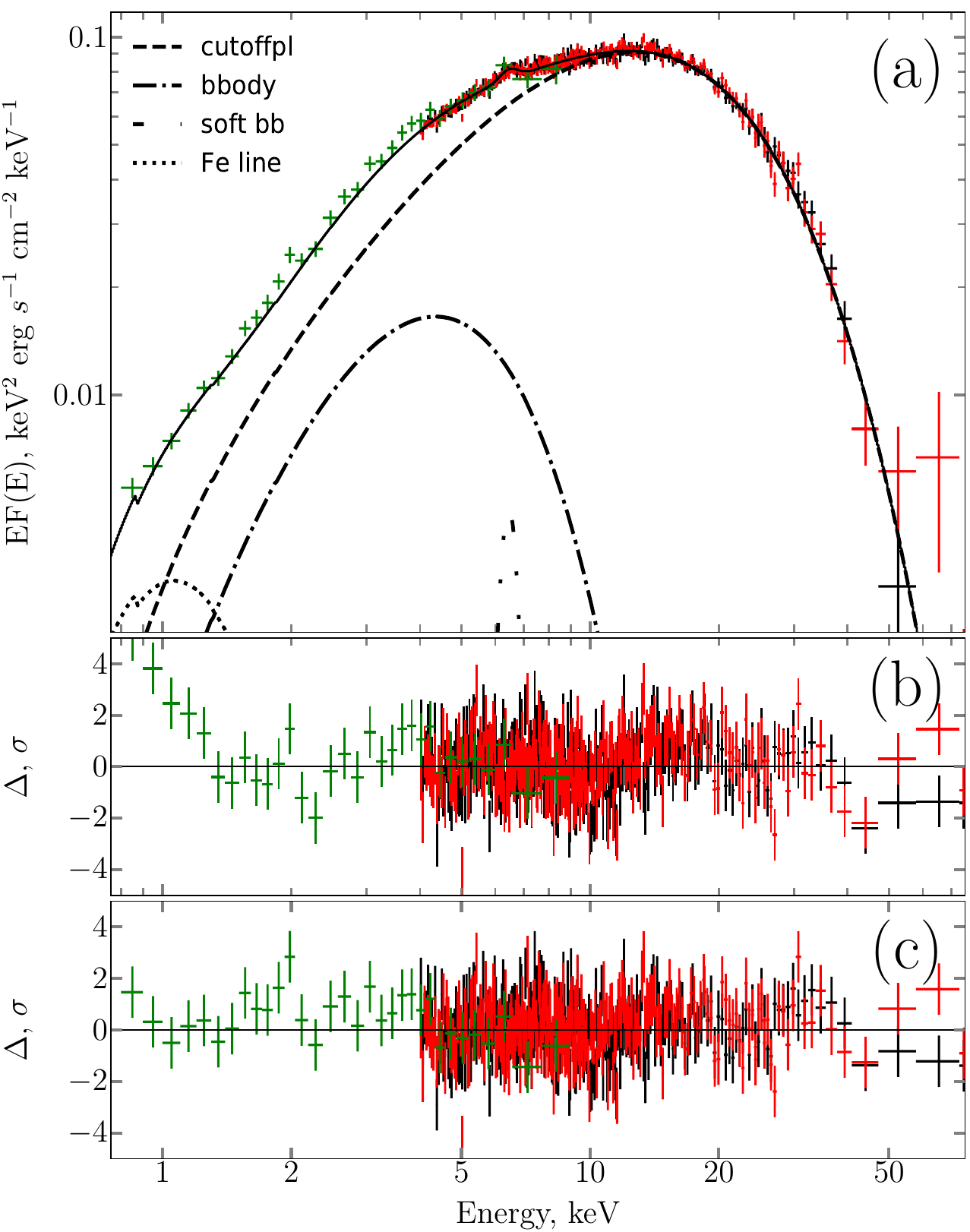}
  \caption{(a) The spectrum of \sxp\ obtained during the first \nustar\ and simultaneous \swiftxrt\ observations and unfolded with {\sc tbabs (cutoffpl + bbody + bbody + gauss)} model.  (b) The residuals for such a model without one of the blackbodies. (c) Residuals for the full model. }
  \label{fig:avespec}
\end{figure}

\begin{table}
\caption{Best-fitting parameters of broad-band spectra of \sxp.}\label{tab:bbparam}
	\centering

 \begin{tabular}{lccc}
     \hline
         \hline
Observation                     & NUOBS02                   &NUOBS04              &NUOBS06\\
    \hline
$N_{\rm H}$, $10^{21}$ cm$^{-2}$    & 2 (fixed)                 & 2 (fixed)          &2 (fixed)\\
$\Gamma$                        & $0.19\pm0.03$             & $0.46\pm0.03$      & $0.32\pm0.05$\\
$E_{\rm cut}$, keV              &$7.00\pm0.18$              &$7.92\pm0.12$       & $7.56\pm0.16$\\
${\rm Norm}_{\rm pl}$, $10^{-3}$ &$5.6\pm0.7$               &$12.6\pm0.7$        & $10.4\pm1.1$ \\ 
$E_{\rm Fe}$, keV                   & $6.41\pm0.07$   & $6.49\pm0.05$  &$6.52\pm0.03$\\
${\rm Norm}_{\rm Fe}$, $10^{-4}$ &$0.87\pm0.11$    &$1.54\pm0.17$   &$1.58\pm0.14$\\
EW$_{\rm Fe}$, eV               & $39\pm6$       & $46\substack{+7 \\ -5}$   & $47\substack{+5 \\ -3} $  \\
$kT_{\rm BB}$, keV              & $1.10\pm0.03$   & $1.26\pm0.02$   &$1.32\pm0.01$\\
$R_{\rm BB}$, km                & $9.2\substack{+0.3 \\ -0.4}$    & $8.5\pm0.4$      &$10.1\pm0.3$\\
$kT_{\rm BB,s}$, keV            & $0.21\pm0.03$   & $0.12\pm0.2$        & $0.21\pm0.03$\\
$R_{\rm BB,s}$, km              & $148\substack{+57 \\ -48}$  & $808\substack{+887 \\ -384}$  & $163\substack{+56 \\ -42}$\\
$C_{\rm FPMB}$                  & $1.016\pm0.003$ & $1.020\pm0.003$ & $1.021\pm0.002$ \\
$C_{\rm XRT}$                   & $1.058\pm0.015$ & $1.000\pm0.012$ & $0.969\pm0.022$ \\
$\chi^{2}_{\rm red}$           &  1.003          & 1.003           & 1.009           \\
d.o.f.                          &         1715    & 1692            & 1669            \\

	\hline
	\end{tabular}
  \label{tab:avespec}

\end{table}

To describe the \sxp\ spectrum in the broad energy range (see
Fig.\,\ref{fig:avespec}a) we used the same models that were considered in
Section\,\ref{sec:nuspec}. It was found that the {\sc comptt} model is unable to
approximate the broadband spectrum and an addition of the blackbody
component is required in the same way as it was required for the {\sc
cutoffpl} model earlier. The performed analysis showed that both combined
models, {\sc cutoffpl + bbody} and {\sc comptt + bbody}, describe all three
broadband spectra equally well, but the {\sc cutoffpl} model has fewer
parameters. Therefore in the following analysis we used the {\sc cutoffpl +
bbody} model as a base one. As before an additional Gaussian
component to describe a prominent emission line at 6.4~keV with the fixed
width at 0.3 keV was added to the model. 

This model approximates the source spectrum quite well, but nevertheless at
the residuals panel a very soft excess at energies $<1$ keV is clearly visible (Fig.\,\ref{fig:avespec}b).
Such an excess was previously detected in
spectra of several XRPs and was proposed to originate from the
material at the inner edge of the accretion disc heated by the central source
\citep{2004ApJ...614..881H}. To describe it we added another
blackbody component, which improved the fit quality
(Fig.\,\ref{fig:avespec}c). 

Results of the approximation of spectra of \sxp\ with the model of {\sc
phabs*(cutofpl+bbody+bbody+gaus)} are presented in Fig.\ref{fig:avespec},
corresponding best-fitting parameters are summarized in Table\,\ref{tab:bbparam}.
It is clearly seen that the model fits well to all spectra and again shows no
systematic deviations which would imply the presence of absorption features.
The `high-temperature' blackbody component with $kT_{\rm BB}\simeq1.1-1.3$ keV is often
registered in spectra of Be-XRBs \citep[see, e.g.,][]{2012A&A...539A..82L,2012MNRAS.421.2407T, 2013MNRAS.436.2054B, 2017ApJ...841...35F} and can be connected with
the emission of the hot parts of the accretion column or hot spots around
these columns, arising by interception of their emission by the
surface of the NS \citep{2013ApJ...777..115P}. 
The typical size of the emission region determined from the data 
agrees well with such an explanation. 

The `soft' blackbody component has a temperature in the range of
0.1--0.2~keV  consistent with the study by \citet{2004ApJ...614..881H}.
Following these authors we can estimate the inner radius of the accretion
disc, assuming that it is responsible for this emission
and has the height-to-radius ratio $H/R=0.1$. For the
first and third \nustar\ observations (where parameters of the soft component
are better constrained) we get an estimation of the inner radius as
$R_{\rm in} \approx 10^{8}$~cm, that corresponds to the  magnetic field
strength $B\approx 0.6\times10^{12}$~G.

\subsection{Cyclotron line search in the broad-band spectrum}

To test finally the hypothesis of a possible presence of a cyclotron
absorption line in the spectra of \sxp, we used an approach proposed by
\citet{2005AstL...31...88T} and recently updated by
\citet{2017AstL...43..175S}. 
The best-fitting model was modified by adding the {\sc gabs} component. After that, the cyclotron line energy $E_{\rm cyc}$ was
varied within the 5--55 keV energy range with the step of 0.5 keV. A corresponding line width was varied with the step of 0.5 keV within the range of 1--6 keV (but smaller than $E_{\rm cyc}$/2).  For each
pair of cyclotron line parameters the position and width were fixed in the
{\sc gabs} model component, the resulting model was used to approximate the
spectrum and the confidence interval for the optical depth of the cyclotron
line was calculated. As a result, none of the combination of the line energy
and its width results in a significant improvement of the fit and
only the upper limit for the optical depth $\tau_{\rm cyc} \lesssim 0.06$
(90 per cent confidence interval) was obtained. This upper limit indicates the
absence of the cyclotron feature in the 5--55~keV energy range, that allows
us to put a limit on the possible strength of the magnetic field on the
surface of the NS in \sxp\ $B < 6\times10^{11}$~G or $B >
6.6\times10^{12}$~G. 

Cyclotron absorption lines often demonstrate variations with the pulse phase \citep[see, e.g.]{2000ApJ...530..429B, 2004A&A...427..975K}, therefore their presence can be more prominent in the phase-resolved spectra.
Considering such possibility, we inspected the phase-resolved spectra, but did not find any additional indication in favour of cyclotron absorption lines.

\section{Conclusions}
\label{sec:concl}

In this work we considered temporal and spectral properties of the Be X-ray pulsar SXP~4.78 in order to estimate the strength of the NS magnetic field in this system.
Despite the previously reported detection of the cyclotron absorption line at $10.3\pm0.2$~keV in the \nustar\ spectrum \citep{2018ATel12234....1A},  we found that the majority of spectral models typically used to fit spectra of XRPs do not require an absorption feature in the 5--55~keV energy band. This implies that the magnetic field is either weaker than $0.6\times10^{12}$~G or stronger than $6.6\times10^{12}$~G.

A few more indirect methods were applied to further constrain the magnetic field strength.
Particularly, the long-term temporal behaviour of the source flux points to the transition of \sxp\ to the propeller regime. An upper limit to the transitional luminosity $\sim4\times10^{35}$\lum\ corresponds to the upper limit on the magnetic field strength $\sim10^{12}$~G.
Another evidence of the relatively weak magnetic field in the source comes from shape of the noise power spectrum. We were able to obtain a lower limit on the break frequency corresponding to the Keplerian frequency at the accretion disc inner edge as $f_{\rm br} > 5$~Hz. This frequency corresponds to the magnetospheric radius around $10^{8}$~cm and, hence, the magnetic field lower than $\sim10^{12}$~G. Similar magnetic field strength was obtained from parameters of the soft blackbody component in the energy spectrum of the pulsar associated with the emission from the inner edge of the accretion disc.

Finally, we were able to discover a millihertz QPO in the power spectrum at $\sim1.6$~mHz, which may originate from the precession of the magnetically wrapped disc \citep{2002ApJ...565.1134S}, providing information about the magnetic field strength.
Using measured QPO frequency we derived the magnetic field strength of $B\sim2\times10^{12}$~G. 
Summarizing all the above, we can conclude that \sxp\ contains a relatively weakly-magnetized NS with the magnetic field strength lower than $10^{12}$~G.

\section*{Acknowledgements}

This work was supported by the grant of the Ministry of Science and High
Education 14.W03.31.0021. We also acknowledge the support from the Academy of Finland travel grants 324550 (SST),  322779 (JP) and 316932 (AAL).
The research has made by using data obtained with
\textsl{NuSTAR}, \textsl{Swift}, \textsl{XMM-Newton}, \textsl{NICER} and \textsl{Chandra} observatories.
Authors are very grateful for Dr. Belinda Wilkes, Director of Chandra X-ray Center,
for approving our ToO observations.
Authors thank {\it Swift} PI, Brad Cenko and Swift SOT team for approving and rapid scheduling of our observations.



\bibliographystyle{mnras}
\bibliography{xte_j0052m723}


%
%


\bsp	
\label{lastpage}
\end{document}